\documentclass{PoS}

\usepackage{float}
\usepackage{subcaption}
\usepackage{graphicx}

\title{Atmospheric muons from electromagnetic cascades}

\ShortTitle{Electromagnetic Muons}

\author{\speaker{Stephan Meighen-Berger}\\
        Technical University Munich\\
        E-mail: \email{stephan.meighen-berger@tum.de}}
\author{Mingyang Li\\
        Technical University Munich\\
        E-mail: \email{limingyang16@mails.ucas.edu.cn}}

\abstract{Atmospheric muons are one of the main backgrounds for current Water- and Ice-Cherenkov neutrino telescopes designed to detect astrophysical neutrinos. The inclusive fluxes of atmospheric muons and neutrinos from hadronic interactions of cosmic rays have been extensively studied with Monte Carlo and cascade equation methods, for example CORSIKA and MCEq. However, the muons that are pair produced in electromagnetic interaction of high energy photons are quantitatively not well understood.
We present new simulation results and assess the model dependencies of the high-energy atmospheric muon flux including those from electromagnetic interactions, using a new numerical electromagnetic cascade equation solver EmCa that can be easily coupled with the hadronic solver MCEq. Both codes are in active development with the particular aim to become part of the next generation CORSIKA 8 air shower simulation package. The combination of EmCa and MCEq accounts for material effects that have not been previously included in most of the available codes. Hence, the influence of these effects on the air showers will also be briefly discussed.}

\FullConference{36th International Cosmic Ray Conference -ICRC2019-\\
		July 24th - August 1st, 2019\\
		Madison, WI, U.S.A.}

\begin{document}

\section{Introduction}
	Atmospheric muons, produced in cosmic ray showers, have been measured by multiple experiments \cite{Aartsen:2015nss, CMS:2010wua, Unger:2003xd}. These measurements usually rely on Monte Carlo simulations, such as CORSIKA \cite{Heck:1998vt, Wentz:2003bp} or FLUKA \cite{Battistoni:2007zza}, for the prediction of the atmospheric particle flux. While such Monte Carlo based codes offer high precision predictions, testing of new interaction models comes with a high computational cost. For this reason cascade equations are more suitable for testing and benchmarking of new models \cite{Gaisser:1990vg, Lipari:1993hd}. Especially for high energy particles, cascade equation methods have a similar precision as Monte Carlo based methods. For these reasons in this paper we use a similar approach as MCEq \cite{Fedynitch:2015zma} for the simulation of electromagnetic cascades. This simulation code, "Electromagnetic-Cascades" (EmCa)\footnote{https://github.com/MeighenBergerS/emca} \cite{Meighen-Berger:2019cxt}, provides three different interaction models, which we use to test the model dependence of atmospheric muons coming from electromagnetic cascades. They differ in their description of pair production and Bremsstrahlung. Currently implemented are the cross sections as calculated by Tsai \cite{Tsai:1973py}, with and without the full screening approximation and those used by EGS (4 and 5) \cite{Nelson:1985ec, Hirayama:2005zm}, which are based on Bethe and Heitler's calculations \cite{Bethe:1934za}. The latter is used in CORSIKA for the simulation of electromagnetic cascades. In Section (\ref{sec:model}) we will give an introduction to the used models and in Section (\ref{sec:MuonProd}) the results for electromagnetic cascades are shown. In the following section results for full shower simulations using MCEq coupled with EmCa are discussed.
	\\
	Due to the differences in the charged atmospheric flux between the models, we also expect a change in the fluorescence. In the final section we will discuss theses changes and their relevance to fluorescence experiments.
	
\section{Models}\label{sec:model}
\begin{figure}[htb]
     	\centering
     	\begin{subfigure}[h]{.49\textwidth}
     	\centering
     	\includegraphics[width=1.\linewidth]{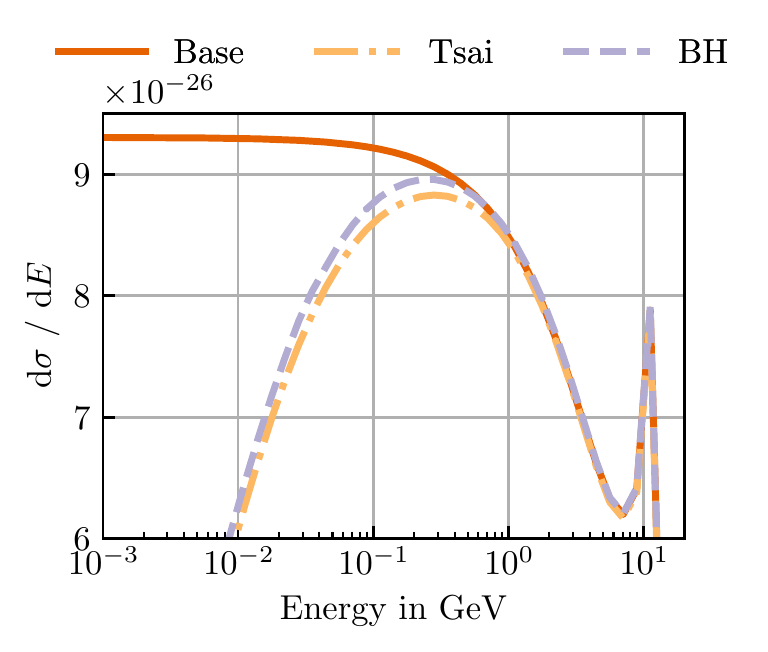}
    \end{subfigure}\hfill%
    \begin{subfigure}[h]{.49\textwidth}
    \centering
    \includegraphics[width=1.\linewidth]{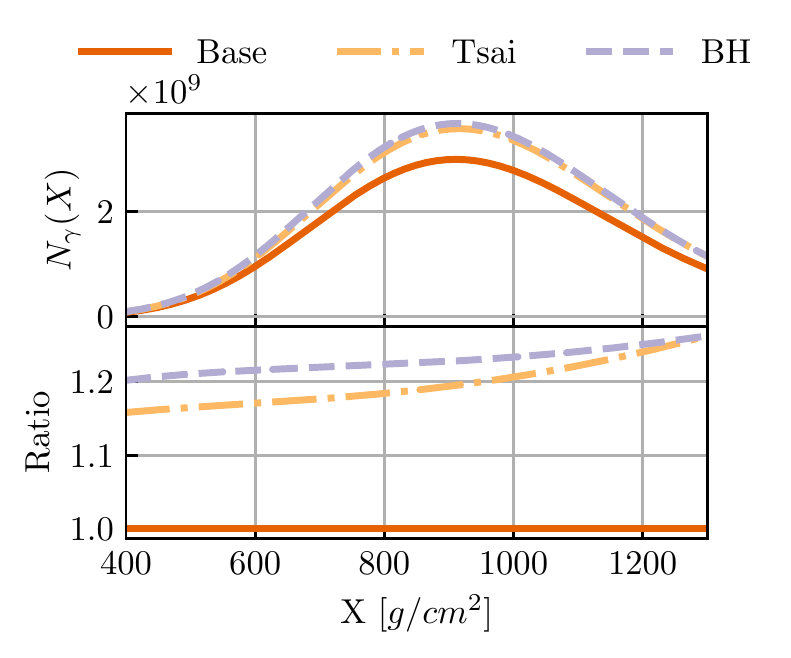}
    \end{subfigure}
    \caption{\textbf{Left:} The differential cross section for pair production for the different models. Orange, solid shows the full screening approximation (Base), light orange dashed-dotted the full Tsai model (Tsai) and in purple dashed the Bethe-Heitler model (BH). The primary photon's energy was set to 1.4 GeV. \textbf{Right:} The number of photons in an atmospheric shower initiated by a primary 10 EeV photon. The color scheme is the same as in the left figure. The bottom plot shows the ratio between the photon numbers using the different models, with Base uses as the reference.}\label{fig:DiffandFlux}
    \end{figure}

\begin{figure}[htb]
     	\centering
     	\begin{subfigure}[h]{.49\textwidth}
     	\centering
     	\includegraphics[width=1.\linewidth]{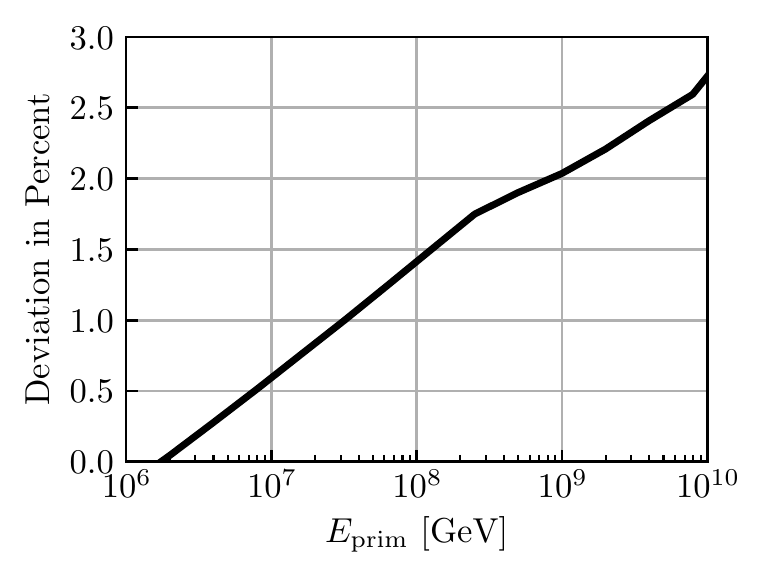}
    	\end{subfigure}\hfill%
    	\begin{subfigure}[h]{.49\textwidth}
    	\centering
    	\includegraphics[width=1.\linewidth]{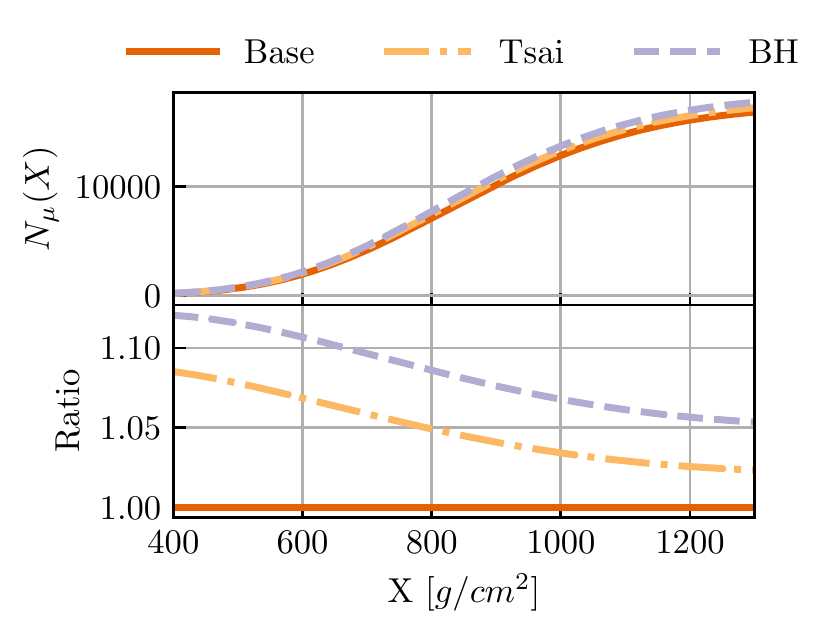}
    	\end{subfigure}
    	\caption{\textbf{Left:} The average difference in photon number across an atmospheric electromagnetic shower between the Tsai and Bethe-Heitler models, with increasing primary photon energy. \textbf{Right:} Depth dependent number of muons in an atmospheric shower initiated by a primary 10 EeV photon. Orange, solid shows the full screening approximation (Base), light orange dashed-dotted the full Tsai model (Tsai) and in purple dashed the Bethe-Heitler model (BH).}\label{fig:EnergyDepMuon}
    \end{figure}

	Tsai's calculations offer a more precise and intrinsic treatment of screening effects, unlike cross sections based on Bethe and Heitler's calculations. This leads to discrepancies in the cross sections. The differential cross sections for pair production using the different models is shown in Figure (\ref{fig:DiffandFlux}). The major difference between the Tsai model using the full screening approximation (Base) and the others, is due to the approximation not being valid anymore for the production of sub GeV particles. The difference between the full Tsai model (Tsai) and the Bethe-Heitler (BH) results are the most relevant for the current discussion. While these discrepancies are minor, the large number of particles and interactions act multiplicative on them. Bremsstrahlung, unlike pair production, has only a very slight screening dependence.
    The difference between the models changes the rate, at which photons interact, increasing the number of photons when accounting for screening. This is shown in Figure (\ref{fig:DiffandFlux}) for an electromagnetic cascade in the atmosphere, initiated by a 10 EeV photon, where we set the cut off energy to 86 MeV. This is the critical energy $E_{\mathrm{crit}}$ according to Rossi \cite{Rossi:1952ph}. Below which collision losses start to dominate the loss of electrons in air. The collision losses, for all models, are modeled using the ESTAR \cite{Berger:2017nist}. The differences between the models shrinks, when increasing the low energy cut off and increases when setting it even lower. The average difference between the full Tsai model and the Bethe-Heitler model is 2.5\% for the depicted shower with a maximal difference of 3.7\%.
    We show the average difference of the photon number for showers initiated by primary photons with different energies in Figure (\ref{fig:EnergyDepMuon}). The increasing difference is expected, since the models differ in the low energy production of particles and with increasing energy more such particles are produced.

\section{Muon Production}\label{sec:MuonProd}
	In electromagnetic cascades muons are produced by photons through pair production. This process is similar to electron pair production, scaled by the mass of the muon. The implemented cross sections are the same as the ones used in CORSIKA defined in \cite{Burkhardt:2002vg}. Since muon production starts to take place at energies where screening is irrelevant one does not need to take its effects into account. Thus the shape of the muon cross sections is far simpler than those of electrons. Additionally we account for radiation losses of mouns by implementing electron pair production by muons as in \cite{Tannenbaum:1990ae} and Bremsstrahlung discussed in \cite{Kelner:1995hu}. Initiating an atmospheric shower with a 10 EeV photon, the number of muons produced in the different models is compared in Figure (\ref{fig:EnergyDepMuon}).
    The average difference, between the Tsai and Bethe-Heitler models, in muon number across the entirety of the shower in Figure (\ref{fig:EnergyDepMuon}) is approximately 3\%. This means the muon number follows a similar trend as the photon number. This difference is now a purely electromagnetic one, as all other cross sections and losses were kept the same across the simulations. This is not accounted for in simulation codes, such as CORSIKA, due to only EGS4 being implemented. The Bethe-Heitler cross section used there is not as precise as the Tsai calculations, meaning CORSIKA simulations of muon fluxes coming from electromagnetic cascades are estimated 3\% too high.
    \subsection{Landau-Pomeranchuk-Migdal Effect}
    	For highly energetic showers, as discussed previously, the Landau-Pomeranchuk-Migdal (LPM) \cite{PhysRev.103.1811, Gerhardt:2010bj} effect has a major impact on the shower development. Accounting for this effect retards the development, due to the cross section at high energies being suppressed. This changes the absolute number of particles in electromagnetic cascades, mostly at low energies. Muons are produced by highly energetic particles, which means their particle number remains unaffected by the LPM effect. This means in the case of muons, the LPM effect can be safely ignored in air showers.
   
\section{Full Shower Simulation}
\begin{figure}[htb]
     	\centering
     	\begin{subfigure}[h]{.49\textwidth}
     	\centering
     	\includegraphics[width=1.\linewidth]{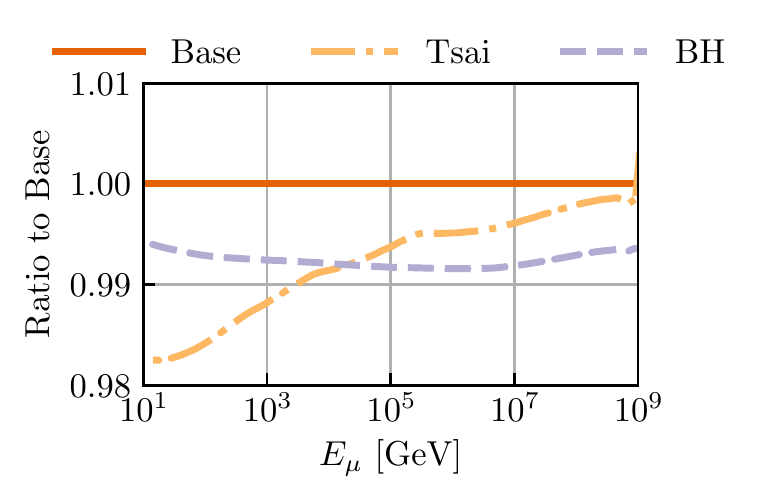}
    	\end{subfigure}\hfill%
    	\begin{subfigure}[h]{.49\textwidth}
    	\centering
    	\includegraphics[width=1.\linewidth]{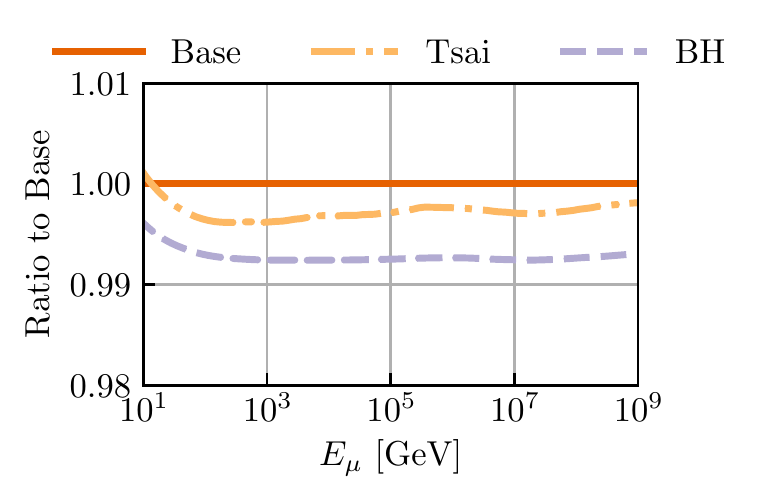}
    	\end{subfigure}
    	\caption{\textbf{Left:} Ratio of the muon flux in an atmospheric shower initiated by a primary 10 EeV proton at sea level. Orange, solid shows the full screening approximation (Base), light orange dashed-dotted the full Tsai model (Tsai) and in purple dashed the Bethe-Heitler model (BH). \textbf{Right:} Ratio of the muon flux when injecting the primary H3a model at sea level. We use the same color scheme as in the left figure.}\label{fig:HadronicMuon}
    \end{figure}

	We extend MCEq to use EmCa to propagate electromagnetic particles. This is done by handing the electromagnetic component MCEq generates to EmCa after each step, which in turn is propagated and then handed back to MCEq. This process is repeated for the entire shower development. For the production of high energy muons this method is sufficient. Should low energy particles be of interest, the EmCa interaction matrices need to be included into MCEq directly for a full shower description. In Figure (\ref{fig:HadronicMuon}) we show a simulation for a single injected 10 EeV proton and the resulting muons binned in energy at sea level. The discrepancies between the models is at maximum 1\%. Note that the main energy regime where the model differences do become relevant, < 1 GeV, can currently not be simulated with MCEq. In Figure (\ref{fig:HadronicMuon}) we also show the results when using the cosmic ray flux defined by the H3a \cite{Gaisser:2013bla} model.
\newpage
\section{Fluorescence}
\begin{figure}[htb]
     	\centering
     	\begin{subfigure}[h]{.49\textwidth}
     	\centering
     	\includegraphics[width=1.\linewidth]{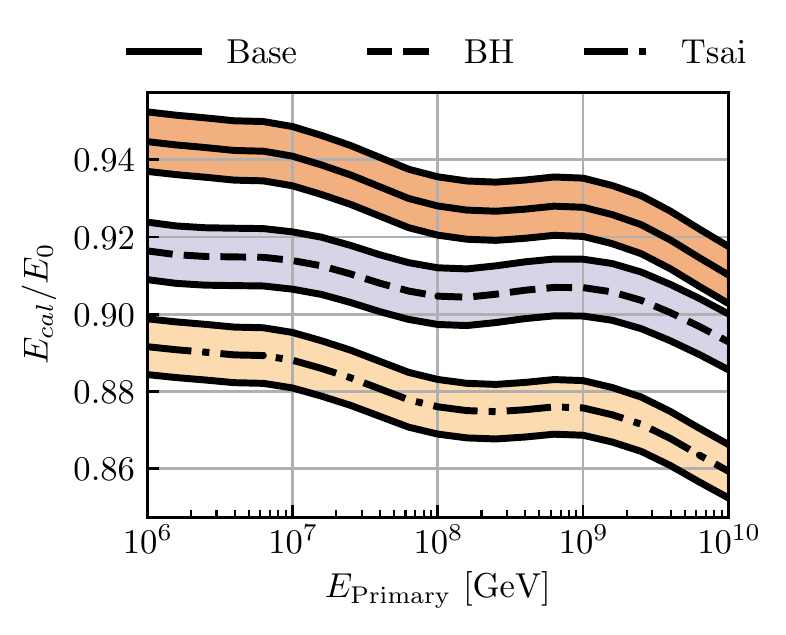}
    	\end{subfigure}\hfill%
    	\begin{subfigure}[h]{.49\textwidth}
    	\centering
    	\includegraphics[width=1.\linewidth]{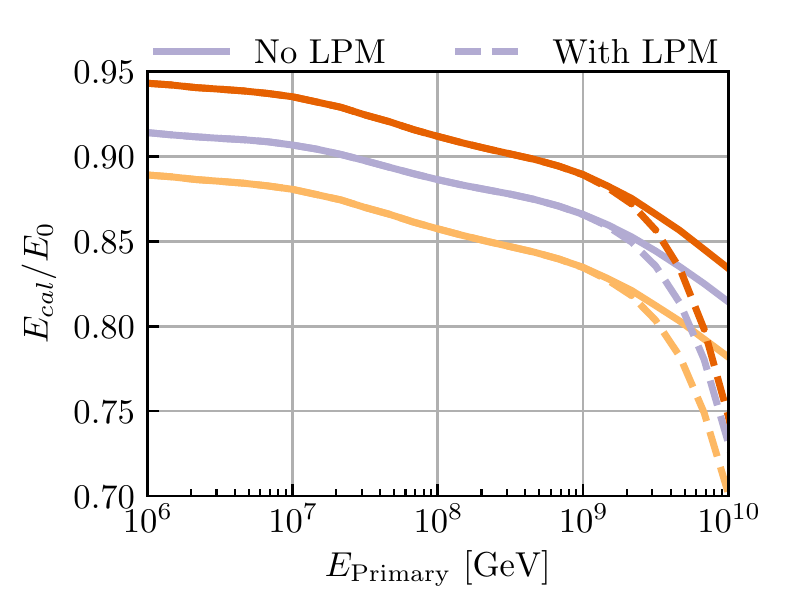}
    	\end{subfigure}
    	\caption{ \textbf{Left:} The deposited energy compared to the injected primary photon energy. Shown are the simulation results for fully developed showers for the different models. We plot the effect a a 1\% shift in $\alpha$ causes using the bands. The different models are plotted as: Solid orange the full screening approximation (Base); Dashed-dotted light orange the full Tsai model (Tsai); Dashed purple the Bethe-Heitler model (BH). The Bethe-Heitler model shows a discrepancy of between 3\% and 4\% with the Tsai model. \textbf{Right:} The deposited energy compared to the primary's energy. Here we use the same color scheme as on the left, while solid plots the results without and dashed with the LPM effect. For these simulations we introduced an observer at sea level.}\label{fig:Fluor}
    \end{figure}

	A natural extension to the previous discussion, is the effect of the electromagnetic models on fluorescence. MCEq cannot be used for an estimate of the fluorescence change, due to the major contribution of low energy charged particles. We run similar simulations again as in Section (\ref{sec:MuonProd}) and estimate the deposited energy, or calometric energy, $E_\mathrm{cal}$, from fluorescence, discussed in \cite{Song:1999wq}. There the deposited energy is estimated as
	\begin{equation}\label{eq:Fluor}
		E_\mathrm{cal} = \alpha\int\limits_0^\infty N_\mathrm{ch}(X)\mathrm{d}X.
	\end{equation}
	In the above equation $N_\mathrm{ch}(X)$ is the number of charged particles at a given slant depth $X$. The slant depth is defined as $X(h) = \int\limits_0^h \rho(l)\mathrm{d}l$. In \cite{Song:1999wq} $\alpha = 2.19$ MeV, which we also use. Figure (\ref{fig:Fluor}) shows the simulation results. Left shows the deposited energy, assuming the shower fully developed, while on the right the case when accounting for the LPM effect and introducing an observer at sea level is plotted.
    The differences between the Bethe-Heitler and Tsai models shown vary between 3\% and 4\%. This means currently, due to the Tsai model being a more exact model, EGS4 and by extension CORSIKA, estimate photon shower energies as too high. Since electromagnetic cascades provide the majority of fluorescent light, due to electrons comprising most of the charged particles in the course of a showers development, these discrepancies will also hold true for hadronic showers. The absolute unaccounted for error this causes in hadronic showers is difficult to quantify and requires further study.
\newpage
\section{Conclusion}
   We show and discuss multiple simulations using EmCa and MCEq to test the effect electromagnetic models have on muon spectra in atmospheric showers. The relevance of screening is shown in Figure (\ref{fig:DiffandFlux}) and that there are differences between the electromagnetic model used in CORSIKA and the more precise model used in EmCa. These differences lead to a primary energy dependent change in the total number of muons across the entire shower development. In the case of electromagnetic showers initiated by a 10 EeV photon these differences averages to 3\%. In the case of hadronic showers, where we were limited to simulation results of particles above 10 GeV by MCEq, the difference is sub 1\% for a 10 EeV initiated shower. The difference become even smaller when injecting a cosmic ray spectrum, such as H3a.
   \\
   Finally we also discussed the change in fluorescence caused by the different models in electromagnetic air showers. Currently, using CORSIKA, energy estimates for electromagnetic cascades and by extension hadronic showers are too high. We calculated the shift for electromagnetic showers to vary between 3\% and 4\%. These effects are especially relevant for fluorescence experiments, for example Auger \cite{Abraham:2009pm} and TA \cite{Tokuno:2012mi}.
   \\
   For both the muon flux and fluorescence the dependence on electromagnetic models in hadronic cascades requires further study.
\newpage
\section{Acknowledgments}
The material presented in this publicationis based upon work supported by the Sonderforschungsbereich Neutrinos and Dark Matter in Astro- and Particle Physics (SFB1258). Furthermore we would like to thank Elisa Resconi, Matthias Huber and Anatoli Fedynitch for the discussions and support.

\bibliography{mybibfile}{}
\bibliographystyle{plain}

\end{document}